\documentclass[aps,prd,a4paper,showpacs]{revtex4}
\usepackage[dvips]{graphicx}

\usepackage{amsmath,amsfonts,amssymb,simplewick}
\begin{document}
\title{\bf A Renormalization Program for Systems with Non-Perturbative Conditions}
\author{S.S. Gousheh}
\email{ss-gousheh@sbu.ac.ir}
\author{F. Charmchi}
\email{f_charmchi@sbu.ac.ir}
\date{\today}
\affiliation{Department of Physics, Shahid Beheshti University G. C., Evin, Tehran 19839, Iran}
\begin{abstract}
In this paper we introduce an alternative renormalization program for 
systems with non-perturbative conditions. 
The non-perturbative conditions that we concentrate on in this paper are confined to be either 
the presence of non-trivial boundary conditions or non-perturbative background fields. 
We show that these non-perturbative conditions have profound effects on all physical 
properties of the system and our renormalization program is consistent with
these conditions. 
We formulate the general renormalization program in the configuration space. 
The differences between the free space  renormalization program and ours 
manifest themselves in the counter-terms as well, which we shall elucidate.
The general expressions that we obtain for the counter-terms reduce to the 
standard results in the free space cases.
We show that the differences between these divergent counter-terms 
are extremely small.
Moreover we argue that the position dependences induced on the parameters of the 
renormalized Lagrangian via the loop corrections, however small, are
direct and natural consequences of the non-perturbative  position dependent 
conditions imposed on the system.
\end{abstract}
\maketitle
\section{Introduction}
\par
The presently successful theory of particle physics is the Standard Model which is a local 
Quantum Field Theory (QFT) including electromagnetic, weak, and strong interactions.
Most problems of physical interest are not exactly solvable and when the problem is amenable 
to perturbation theory and describable by a renormalizable local QFT, the calculations are usually 
done perturbatively. 
The renormalization program for these theories are defined to remove the divergences order by order in 
perturbation theory, by redefining the physical parameters of the system.
Most of the renormalization programs presented up to now, as far as we know, have been for systems 
which are not subject to any non-perturbative conditions, such as the presence of non-trivial boundary 
conditions or non-perturbative background fields. The appropriate renormalization program for these 
systems is the usual free space renormalization program.
\par
The renormalization program started in Quantum Electrodynamics (QED).
The initial successes of QED were based upon the renormalization program 
initiated by Tomonaga\,\cite{Tomonaga46, Tomonaga47, Tomonaga3, Tomonaga4}, 
Schwinger\,\cite{Schwinger1, Schwinger2, Schwinger3, Schwinger4} and Feynman\,\cite{Feynman1, Feynman2, Feynman3} 
and developed to a general form by Dyson\,\cite{Dyson}. 
The amazing success of this free-space renormalization program to predict the radiative corrections for 
various quantities in QED, such as the electron $g$-factor, has left little doubt on the validity of this program.
Analogous free space renormalization programs have been an essential and integral part of the development 
of the Standard Model. It is extremely important to note that all of these QFT calculations including their 
renormalization programs are perturbative calculations based on the free space.
Before discussing our renormalization program we like to emphasize that the case of ($1+1$) dimensions is very special in the sense that
the renormalization program is equivalent to normal ordering \,\cite{Coleman}.
For a comprehensive review of free space renormalization program and its history  see for example\,\cite{Weinberg}. 
\par
The main issue that we discuss in this paper is that the presence of non-perturbative conditions 
in the system, such as non-trivial boundary conditions, non-perturbative background fields such 
as solitons, and non-trivial background metrics have profound effects on all of the physical aspects 
of the system which cannot be taken into account perturbatively.
The boundary conditions that we consider include the phenomenological ones, such as the the ones 
appropriate for the bag model of nucleons which effectively provide the confinement mechanism of the 
low energy QCD\,\cite{BagModel1,BagModel2,BagModel3,BagModel4,BagModel5}.
The appropriate renormalization program should be self-contained, and automatically take into account 
the aforementioned conditions in a self-consistent manner. Moreover, the renormalization program should 
not only be consistent with these non-perturbative conditions, but also should emerge naturally from 
the standard procedures. The presence of these non-perturbative conditions breaks the translational symmetry 
of the system, which obviously could have many profound manifestations. 
In particular the linear momentum will no longer be a conserved quantity. 
For these systems, the use of free space renormalization program in which the momentum appears to all 
orders in perturbation theory, is certainly not appropriate in momentum space and might not be appropriate in general. 
We should mention that recently much work has been done on the renormalization program for
the system within the context of non-commutative geometry, and these fall with the category 
of non-trivial renormalization program\,\cite{Non-Commu1,Non-Commu2,Non-Commu3}.
\par
Most importantly, the information about the non-perturbative conditions is carried by the full set 
of the $n$-point functions. We expect and shall show that the breaking of the translational symmetry 
could force all of the $n$-point functions of the theory to have, in general, non-trivial position dependence
in the coordinate representation. 
This occurs with certainty for the case of non-trivial boundary conditions.
For the case of non-trivial background fields this occurs only in non-perturbative cases where the 
Green\rq{}s functions are altered.
In this paper we concentrate on these two cases.
The case of non-trivial background metrics in finite volume has already been investigated 
\,\cite{Balaban1,Balaban2,Balaban3}, and position dependent counter-terms have been obtained.
The procedure to deduce the counter-terms from the $n$-point functions in a renormalizable perturbation 
theory is standard and has been available for over half a century.
In our renormalization program, this will  
lead to a set of uniquely defined position dependent counter-terms. 
However, as we shall explicitly show, the differences between the two divergent counter-terms, i.e. 
the position dependent and the free ones, are generically not only finite but also extremely small. 
\par
In the absence of any non-perturbative conditions, the momentum space and configuration 
space renormalization programs are equivalent and can be used interchangeably\,\cite{Collins}. 
In this paper we set up a general formalism for the configuration space renormalization programs 
which can be used even when the non-perturbative conditions are present.
In the process of removing the divergences, the counter-terms are determined self-consistently and unambiguously 
by the standard procedure itself. These counter-terms turn out to be position dependent, as a direct consequence of 
position dependence of the $n$-point functions. 
In section \ref{section 2} we set up a renormalization program for the problems with non-perturbative conditions, 
and present four sets of reasonings in favor of our renormalization program.
In section \ref{section 3} we summarize our results.
In the appendix we illustrate our results our results using a very simple model.
\section{Renormalization program for problems with Non-perturbative conditions}\label{section 2}
Our starting point is the standard expression for the $n$-point functions 
\begin{align}\label{npf}
\langle\Omega|T\{\phi(x_1)\ldots\phi(x_n)\}|\Omega\rangle=
\lim\limits_{T\rightarrow\infty(1-i\epsilon)}\frac{\langle0|T\{\phi_{I}(x_1)
   \ldots\phi_{I}(x_n)~\textrm{e}^{-i\int^{T}_{-T}\textrm{d}^4x~\mathcal{H}_{I}}\}|0\rangle}
  {\langle0|T\{\textrm{e}^{-i\int^{T}_{-T}\textrm{d}^4x~\mathcal{H}_{I}}\}|0\rangle}.
\end{align}
Since most problems of physical interest are not exactly solvable, when applicable, one usually resorts 
to perturbation theory by expanding the exponential, and uses the interaction picture for convenience.
This expansion in principle contains infinite number of terms all of which are propagators defined by  
\begin{align}
 G(x,y)=\langle\Omega|T\{\phi(x)\phi(y)\}|\Omega\rangle=\contraction{}{\phi}{(x_1)}{\phi}
                                                  \phi{(x)}\phi{(y)}.
\end{align}
The modes appearing in the expansion of $\phi(x)$ must be chosen 
to be the eigen-modes of the system. These modes and the resulting 
Green\rq{}s functions must satisfy the following differential equations, 
in addition to the boundary conditions imposed on the system,  
\begin{align}
 D_x\phi(x)=\omega\phi(x)
 ,\quad D_xG(x,y)=-i\delta(x-y), 
\end{align}
where the  differential operator $D_x$ directly emerges from the Euler-Lagrange equation for the system.
In order to show the procedure, we need to be more concrete and therefore use the $\lambda\phi^4$ theory as a 
generic example. The Lagrangian density for a real scalar field with $\lambda\phi^4$ self-interaction suitable 
for the trivial case or the case with non-trivial (NT) boundary condition is
\begin{align}\label{L}
\mathcal{L}=\frac{1}{2} [\partial_\mu\phi(x)]^2-\frac{\lambda_{0}}{4!}\big[\phi(x)^2+\frac{6m_{0}^2}{\lambda_{0}}\big]^2,
\end{align}
where $m_0$ and $\lambda_0$ are the bare mass and bare coupling constant, respectively.
For the case with a solitonic background one should replace $m_0^2\rightarrow-m_0^2$.   
After rescaling the field by $\phi=Z^\frac{1}{2}\phi_r$ and utilizing the standard 
procedure for setting up the renormalized perturbation theory, the Lagrangian becomes
\begin{align}\label{r.L}
  \mathcal{L}=\frac{1}{2} [\partial_\mu\phi_r]^2-\frac{1}{2} m^2\phi_r^2-
    \frac{\lambda}{4!}\phi_r^4-\frac{3}{2}\frac{m_0^4}{\lambda_0}
+\frac{1}{2}\delta_z [\partial_\mu\phi_r]^2-\frac{1}{2}\delta_m\phi_r^2-\frac{\delta_\lambda}{4!}\phi_r^4,
\end{align}
where $\delta_m$, $\delta_\lambda$ and $\delta_z$ are the counter-terms, and $m$ 
and $\lambda$ are the physical mass and physical coupling constant, respectively.
The relationship between the bare and physical quantities are
\begin{align}\label{c}
\delta_z=Z-1,~~~\delta_\lambda=\lambda_0 Z^2-\lambda\qquad\textrm{and}\qquad\delta_m=m_0^2 Z-m^2.
\end{align}
Notice that we can draw an important condition from Eq.\,(\ref{c}) that any position 
dependence induced in the counter-terms in the renormalization program, will necessarily 
induce an opposite position dependence in the parameters of the theory.
For example for the mass we conclude 
\begin{align}
m^{2}_{}=\frac{\delta_m^{\textrm{NT}}+m^{2}_{\textrm{NT}}}{Z_{\textrm{NT}}}
=\frac{\delta_m^{\textrm{free}}+m^{2}_{\textrm{free}}}{Z_{\textrm{free}}}
\end{align}
since the free case has no position dependence, neither should the combination of 
non-trivial quantities shown.
\par
It is extremely important at this point to distinguish between three separate cases.
First, the $\lambda\phi^4$ theory in the topologically trivial sector in free space, 
which we shall refer to as the ``free'' case.
Second, the $\lambda\phi^4$ theory in the topologically trivial sector with non-trivial 
boundary conditions imposed, which we shall refer to as the ``non-trivial boundary'' case.
Third , the $\lambda\phi^4$ theory in the topologically non-trivial sector,  
which we shall refer to as the ``soliton'' case.
Most of the material presented from this point on is common between these three cases.
However, there are delicate differences which we shall highlight at appropriate points.
\par
In the context of renormalized perturbation theory, as indicated in Eq.\,(\ref{r.L}), 
we can symbolically represent the first few terms of the perturbation expansion of 
Eq.\,(\ref{npf}). The results for the two-point and four-point functions up to 2-loops 
in $\lambda\phi^4$ theory are shown in Eq.\,(\ref{diag1}) and Eq.\,(\ref{diag2}).
In Eq.\,(\ref{diag2}) we have only shown the diagrams of the $s$-channel for simplicity.
In these diagrams the counter-terms should eliminate the corresponding divergences. 
The first argument in favor of our program is as follows.
\begin{align}
   \raisebox{-6mm}{\includegraphics[width=12.7cm]{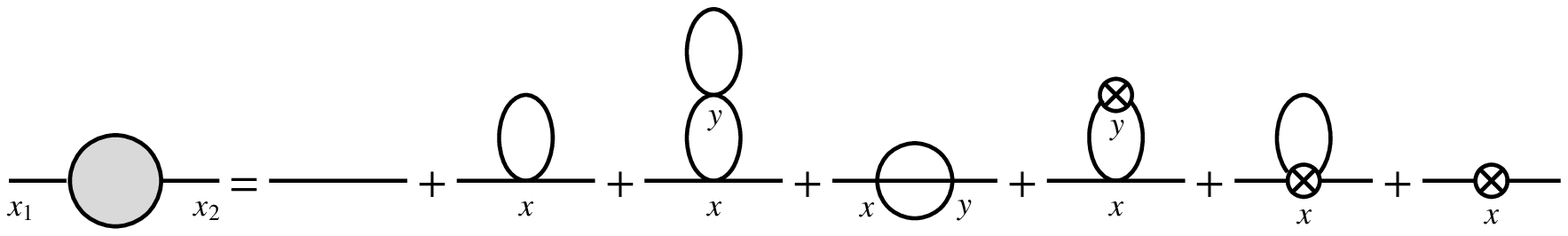}}.
\label{diag1}\\
   \raisebox{-6mm}{\includegraphics[width=12.7cm]{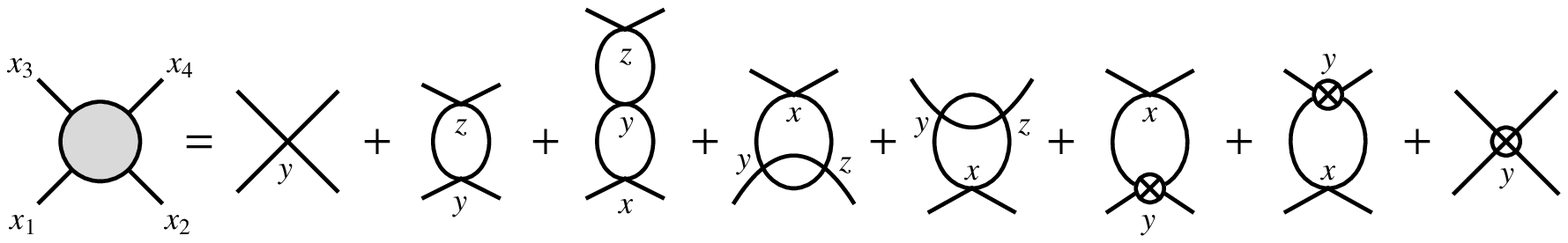}}.\label{diag2}
\end{align}
The propagators appearing in these diagrammatic expansions are the results of the
contractions of the fields in Eq.\,(\ref{npf}) after the perturbation expansion of the 
exponential. 
Therefore, all of the propagators appearing in these perturbation expansions 
should be of the same kind, i.e. the free ones or the non-trivial ones. 
The form of the propagator appropriate for each case is 
dictated by the physical conditions which determine the problem. The propagators appropriate 
for the problem directly appear in the renormalization procedure and thus both are fixed by
the nature of the problem. As we shall show, use of the free space renormalization program 
for the non-trivial cases requires using free propagators for all of the internal lines, 
and this might lead to inconsistencies.
\par
We should note that the diagrammatic expansion illustrated in Eq.\,(\ref{diag1}) and Eq.\,(\ref{diag2}) 
could be representations of either one of the three cases mentioned above. The difference lies only in the 
propagators. For implementing the renormalization program for the $n$-point functions, one must first study 
the Green\rq{}s functions in the free and non-trivial systems. To calculate the counter-terms we need the 
divergent parts of the $n$-point functions. In the free $\lambda\phi^4$ case, obviously the position and 
momentum space renormalization programs are equivalent. In this case these divergences in the momentum 
space renormalization program are due to integration over large momenta in the loops of the Feynman diagrams. 
The integration over large momenta corresponds to integration over infinitesimal distances in the coordinate 
space, i.e. when any of two internal points are close to each other.
\par
Our second argument in favor of our program has two parts: first we show that 
a necessary requirement for self-consistency of the renormalization program is that
all of the internal propagators have to be of the same form.
Second we show that in the process of connecting the internal and external propagators
through the internal points immediately adjacent to the external ones, the requirement
of consistency mandates that all of the propagators have to be of the same form.
Since the external propagators by definition have to be consistent with the non-perturbative
conditions on the system, we conclude that {\em all} of the propagators and the resulting 
counter-terms have to be of such form.
\par
we start by studying the two-point function. Obviously the Green\rq{}s function in the non-perturbative 
cases will have non-trivial position dependence in the coordinate representation. These position 
dependences do not disappear when the end-points approach each other. We start by considering 
the one-loop correction to the two-point function. Consider the first, second and the last diagram 
on the r.h.s. of Eq.\,(\ref{diag1}). Our renormalization condition is identical to the usual ones, which 
states that the exact propagator close to its pole should be equal to the propagator represented by 
the first term. This implies that the second term and the counter-term should cancel each other. That is
\begin{equation}
 \int\textrm{d}^4 x~G(x_1,x)\left\{\frac{\lambda}{2}G(x,x)+\delta^{(1)}_m\right\}G(x,x_2)=0.  
\end{equation}
Since $x_1$ and $x_2$ are arbitrary, from this integral expression one concludes
 that the expression for the counter-term to one-loop is \,\cite{Gousheh-2},  
\begin{equation}\label{m.c}
   \delta^{(1)}_m(x)=\frac{-i}{2}\raisebox{-2.7mm}{\includegraphics[width=1.3cm]{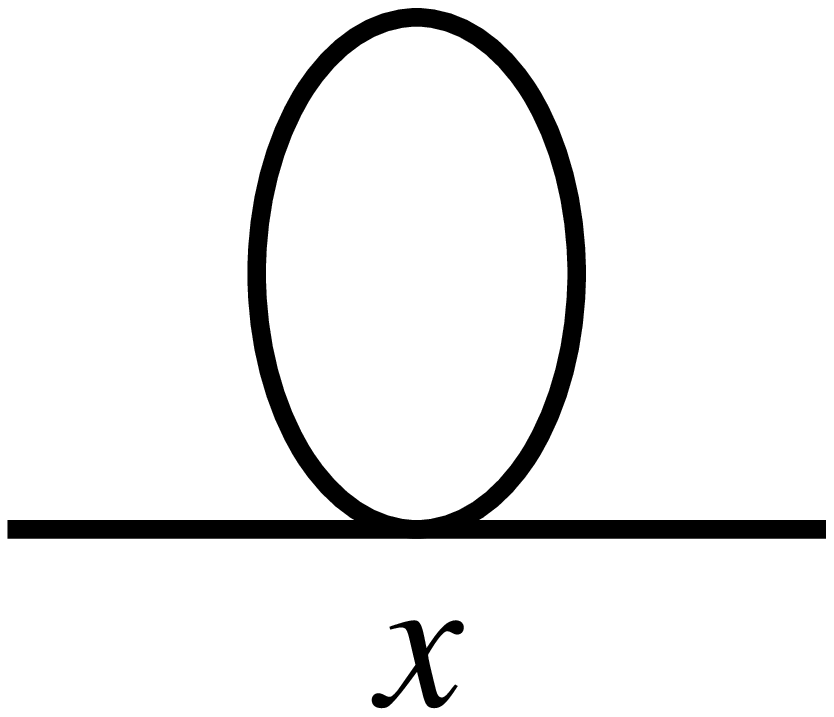}}
   =\frac{-\lambda}{2}G(x,x).
\end{equation}
This is the general desired result for the trivial or non-trivial cases, 
i.e. $G(x,y)$ could be the propagator of the real scalar field for any 
of those cases. From this expression we conclude that the counter-terms 
in the non-trivial cases have non-trivial position dependences. 
In the trivial case our expression for $\delta_{m}^{(1)}$ reduces to the standard result,
\begin{align}\label{1-loop mass counter-term}
\delta_{m,\textrm{free}}^{(1)}
=-\frac{\lambda}{2}\lim_{y\rightarrow x}\int\frac{\textrm{d}^{d}k}{(2\pi)^{d}}\frac{ie^{i(x-y)}}{k^2-m^2+i\epsilon}
=-\frac{\lambda}{2}\int\frac{\textrm{d}^{d}k}{(2\pi)^{d}}\frac{i}{k^2-m^2+i\epsilon}
=-\frac{\lambda}{2(4\pi)^{\frac{d}{2}}}\frac{\Gamma(1-\frac{d}{2})}{(m^2)^{1-\frac{d}{2}}}. 
\end{align}
\par
Next we consider the one loop contribution to the four-point function and the calculation 
of the first order correction to the vertex counter-term. We show the calculations only for 
the $s$-channel, since the $t$ and $u$ channels are calculated similarly. For this purpose 
consider the first, second and the last diagrams on the r.h.s. of Eq.\,(\ref{diag2}).
As usual the renormalization condition is that the divergent part of the second term and the counter-term 
should cancel each other. That is,
\begin{align}\label{s.channel.diagrams}
\lim_{z\rightarrow y}\int\textrm{d}^{4}y\int\textrm{d}^{4}z~G(x_1,y)G(x_2,y)
\left\{\frac{(-i\lambda)^2}{2}G^2(y,z)-i\delta_{\lambda}^{(2)s}(y)\delta^{4}(y-z)\right\} 
  G(z,x_3)G(z,x_4)=0,
\end{align}
where the limit $z \rightarrow y$ has been implemented for the last two propagators and 
the superscript $s$ stands for the $s$-channel. Since $x_1,\dots ,x_4$ are arbitrary, one concludes 
from this integral expression that the expression for the counter-term to one-loop should be
\begin{align}\label{v.c}
 \delta^{(2)}_{\lambda}(y)=\lim_{z\rightarrow y}\frac{-i}{2}\left(~~\raisebox{-5.0mm}{\includegraphics[width=4.5cm]{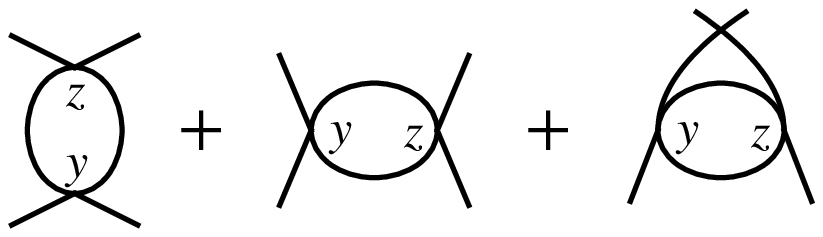}}\right)
  =\lim_{z\rightarrow y}\frac{3i\lambda^2}{2}\int\textrm{d}^{4}z~G^2(y,z),
\end{align}
which in the trivial case reduces to the standard result expression,
\begin{align}
\delta_{\lambda,\textrm{free}}^{(2)}=\frac{3i\lambda^2}{2}\lim_{z\rightarrow y}
\int\textrm{d}^{d}z\int\frac{\textrm{d}^{d}k_1}{(2\pi)^{d}}\frac{\textrm{d}^{d}k_2}{(2\pi)^{d}} 
\frac{ie^{ik_1.(y-z)}}{k_{1}^{2}-m^2+i\epsilon}~\frac{ie^{ik_2.(y-z)}}{k_{2}^{2}-m^2+i\epsilon}
=\frac{3\lambda^2}{2(4\pi)^{\frac{d}{2}}}\frac{\Gamma(2-\frac{d}{2})}{(m^2)^{2-\frac{d}{2}}}. 
\end{align}
\par
To prove the first piece of our second argument we need to concentrate on higher order corrections.
First consider the second order correction to the two-point function. It is easy to show that the divergent parts 
of the third and fifth diagram on the r.h.s. of Eq.\,(\ref{diag1}) cancel each other exactly, provided that the 
counter-term in the fifth diagram is chosen in accordance with Eq.\,(\ref{m.c}), and the propagator in upper 
loop of the third diagram is of the same form as in the first order diagram shown in Eq.\,(\ref{m.c}).
This can be easily seen by combining the divergent parts of those two diagrams as follows, 
\begin{align}\label{3and5}
 \frac{(-i\lambda)}{2}\int\textrm{d}^4 x~\textrm{d}^4 y~G(x_1,x)~G^{2}(x,y)
\left\{\frac{(-i\lambda)}{2}~\textrm{pole}\{G(y,y)\}-i\delta_m (y)\right\}G(x,x_2)=0.
\end{align}
The fourth and sixth diagram in Eq.\,(\ref{diag1}) both have divergences of order 
$1/\epsilon$ and $1/\epsilon^2$ where $\epsilon=4-d$. The $1/\epsilon$ terms 
of these two diagrams should cancel each other, in order for the first order result 
Eq.\,(\ref{m.c}) to hold true. For this to happen, the internal propagators in the 
loop diagram in Eq.\,(\ref{v.c}), which determine the first order vertex counter-term, 
should be of the same form as the propagators appearing in the fourth diagram of the 
two-point function. Then the requirement of the cancellation of $1/\epsilon^2$ terms 
automatically determines the second order part of the mass counter-term, which we shall 
derive and present below. 
\par
Analogous arguments can be presented for the four-point function given in Eq.\,(\ref{diag2}).
Consider the third, sixth and seventh diagrams all in the $s$-channel. 
The $1/\epsilon$ divergences of these diagrams should cancel each other, in order for the 
$O(\lambda^2)$ divergence in the vertex counter-terms given by Eq.\,(\ref{v.c}) to hold true. 
Therefore, the internal propagators shown in that equation should be of the same form as those 
of the second diagram on the r.h.s. in Eq.\,(\ref{diag2}).
Moreover, the divergences resulting from the collapse of the lower loop in the fourth and 
the upper loop in the fifth diagram should be canceled by the sixth and seventh diagrams 
(when the $t+u$ parts of the vertex counter-terms are used), respectively. 
This would happen if and only if the $O(\lambda^2)$ divergence of 
these counter-terms are in accord with Eq.\,(\ref{v.c}).
This in turn implies that the mentioned propagators in the loops in the fourth and fifth 
diagrams should be of the same form as those displayed in Eq.\,(\ref{v.c}).
In that case the upper two propagators in the fourth diagram and lower two of the fifth diagram should also be of the 
same form as the propagators shown in the sixth and seventh diagrams in the corresponding channels. Obviously if the 
internal propagators of the fourth and fifth diagram are not of the same kind one would encounter inconsistencies. 
Next we can consider diagrams of order $\lambda^3$ and above. 
In particular, consider diagrams with either an extra closed propagator 
or a mass counter-term inserted on any of 
its internal propagators. Obviously the divergent parts of these terms should cancel each other, and this occurs only if the mass counter-term 
is chosen according to Eq.\,(\ref{m.c}). The above arguments clearly show that all of internal propagators should be of 
the same kind.  This proves the first part of our second argument. 
\par
Now we are in a good position to determine the second order mass and field  strength counter-terms 
(to first order $\delta_z=0$). Now consider the fourth, sixth and seventh diagrams on the r.h.s. of 
Eq.(\,\ref{diag1}). Using Eq.(\,\ref{v.c}), we obtain  
\begin{align}\label{2.loop.mass}
 &\int\textrm{d}^4 x~G(x_1,x)\nonumber \\
 &\times\lim_{y\rightarrow x}\bigg\{\lambda^2\int\textrm{d}^4 y\bigg[
 \bigg(\frac{3}{4}G(x,x)G^2(x,y)-\frac{1}{6}G^3(x,y)\bigg)G(x,x_2)
 -\frac{1}{12} (y-x)^\mu(y-x)^\nu G^{3}(x,y){\frac{\partial^2 G(y,x_2)}{\partial y^\mu\partial y^\nu}}\mid_{y=x}\bigg]\nonumber \\
 &+\bigg({\overleftarrow{\frac{\partial}{\partial x^\mu}}} i{\delta}^{(2)}_z(x){\overrightarrow{\frac{\partial}{\partial x_\mu}}}
 -i\delta^{(2)}_m(x)\bigg)G(x,x_2)\bigg\}=0. 
\end{align}
From this integral expression we find the second order mass counter-term, 
\begin{align}\label{mass-counter-term} 
 \delta^{(2)}_m(x)=\lim_{y\rightarrow x}\frac{i\lambda^2}{6}\int\textrm{d}^4 y\left\{G^{3}(x,y)-\frac{9}{2}G(x,x)G^2(x,y)\right\}.
\end{align}
However, for the field strength counter-term the integral of the internal and external 
propagators are entangled in such a way that it is in general difficult to extract and isolate 
the infinite part of the integral which would yield the $\delta^{(2)}_z(x)$ counter-term. 
We can simplify the expression involving $\delta^{(2)}_z(x)$ by using the coordinate 
transformations $x=x_0-\frac{u}{2}$ and $y=x_0+\frac{u}{2}$ to symmetrize the second order 
derivative term in Eq.\,(\ref{2.loop.mass}) to obtain,  
\begin{align}\label{z-counter-term}
\int\textrm{d}^4 x_0 ~\frac{\partial G(x_1,x_0)}{\partial {x_0}^\mu}\bigg\{i\delta^{(2)}_z(x_0)\frac{\partial G(x_0,x_2)}{\partial {x_0}_\mu}+
\lim_{u\rightarrow 0}\frac{\lambda^2}{12}\int\textrm{d}^4 u~u^\mu u^\nu G^{3}(x_0-\frac{u}{2},x_0+\frac{u}{2})
 \frac{\partial G(x_0,x_2)}{\partial {x_0}^\nu}\bigg\} =0.
\end{align}
We obtain following expression involving $\delta^{(2)}_z(x)$,
\begin{align}
 \eta^{\mu\nu}\delta^{(2)}_z(x)=i
         \lim_{u\rightarrow 0}\frac{\lambda^2}{12}
         \int\textrm{d}^4 u~u^\mu u^\nu G^{3}(x-\frac{u}{2},x+\frac{u}{2}),
\end{align}
which in the trivial and massless case reduces to the standard expression,
\begin{align}
\delta_{z,\textrm{free}}^{(2)}=\frac{i\lambda^2}{48}
\int\textrm{d}^{4}u~u^2\left[\int\frac{\textrm{d}^{4}u}{(2\pi)^4}\frac{ie^{ik.u}}{k^2-m^2+i\epsilon}\right]^{3}
~\stackrel{{\tiny{d=4-\epsilon}}}{\longrightarrow}
~\delta_{z,\textrm{free}}^{(2)}=\frac{\lambda^2}{12(4\pi)^4}(-\frac{1}{\epsilon}).
\end{align}
We expect the higher order contributions to $\delta_z(x)$ to be even harder to isolate, if not impossible.
The expressions that we have obtained for $\delta^{(2)}_m(x)$ and $\delta^{(2)}_z(x)$ 
in Eq.(\,\ref{mass-counter-term}) and Eq.(\,\ref{z-counter-term}), also indicate that the  
consistency of the renormalization program mandates that {\em all} of the {\em internal} 
propagators to be of the same kind.
\par
Up to now we have only shown that a necessary condition for the consistency of renormalization 
program is that all of the internal propagators to be of the same kind, either trivial or non-trivial. 
Now, we show that this condition is insufficient. This is accomplished by showing that in order to 
connect the internal and external propagators, {\em all} of the propagators should be of the same form.
The argument is as follows. 
The external propagators connect the external points to their adjacent internal points.
Obviously these propagators should satisfy the same set of conditions on {\em all} of
these points.
If we insist that the propagators satisfy the non-trivial conditions imposed on the system
at the external points, they should satisfy the same conditions on the internal ones. 
This starts a cascade process forcing all of the propagators to be of the same form. 
Even if one is willing to consider exotic propagators resulting from the contraction of~ 
$
\contraction{}{\phi}{\,_{\textrm{trivial}}}{\phi}
\phi_\textrm{trivial}\phi_\textrm{non-trivial}
$
, one would run into mathematical difficulty of contracting 
fields of different nature and possibly with different number 
of degrees of freedom (e.g. $\aleph_0$ and $\aleph_1$).
This proves our second set of reasoning.
\par
The expression for $\delta _z$ provides us with the third set of reasonings in favor of our renormalization program.
Since the expressions for the counter-terms in higher orders are extremely convoluted and entangled,
e.g. the expression for $\delta^{(2)}_z$, one  does not have the freedom to choose the counter-terms 
arbitrarily. 
The forms of the counter-terms have to be consistent with the Green\rq{}s functions appropriate 
to the problem, or else the renormalization program's self consistency might be compromised.
\par
The fourth reason in favor of our program is that the loss of translational symmetry in problems 
with non-trivial conditions in general requires the cancellation of all of the divergences to occur locally. 
That is, the divergences will not have in general a constant value throughout the space.
This implies that the counter-terms should be in general position dependent.
\section{Conclusion}\label{section 3}
In this paper we have shown that a consistent renormalization program for the problems with non-perturbative conditions
can be formulated in the configuration space. 
We have presented four sets of reasonings in favor of our renormalization program.
In our program all of the propagators appearing in the perturbative 
expansion of the $n$-point functions should be consistent with the non-perturbative conditions imposed on the
system. This in turn implies that the radiative corrections to all of the input parameters of
the theory, including the mass, will be position dependent. 
As we shall show in the appendix, the induced position dependences on the counter-terms are extremely small.
We believe that the position dependences induced in the parameters of the Lagrangian, including the 
mass and coupling constants, via the loop corrections, however small, are of fundamental importance.
They are a direct and natural manifestation of the non-perturbative  position dependent conditions imposed
on the system.
It is important to note that usually the free space renormalization program also works for low orders 
in perturbation theory, in the sense that it eliminate 
divergences and render the processes within the problem computable.
However, the difference between the two programs leads to small finite differences in the final results.   
We have explored some of the consequences of this renormalization program in connection with
the NLO Casimir energy \cite{Gousheh-2, Gousheh-3}, and radiative correction to the mass of the soliton 
\cite{Gousheh-4}. 
\section*{Acknowledgment}
We would like to thank the research office of the Shahid Beheshti University for financial support. 

\appendix
\section{A simple illustrative example for the comparison between trivial and non-trivial counter-terms}
In this appendix we compute the mass counter-term for a simple $\lambda\phi^4$ model in $1+1$ dimensions with 
Dirichlet boundary conditions (BC).
We then compare this with the analogous counter-term without any boundary conditions.
The Green\rq{}s function for the case with the boundary conditions imposed is,
\begin{center}
\begin{figure}\hspace{0.0cm}
  \includegraphics[width=9.cm]{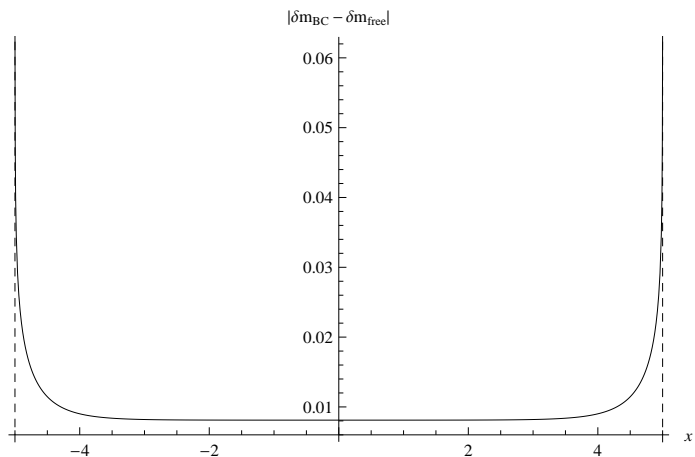}
 \caption{The difference between the mass counter-terms of a scalar field confined with Dirichlet boundary
                condition and the free case in $1+1$ dimensions, when $m=1$, $\lambda=0.1~(m^2)$ and $a=10~(m^{-1})$.
                Note that the difference between these two infinite quantities is significant only very close to the 
                boundaries. As $a\to\infty$ the difference goes to zero for $-\frac{a}{2}<x<\frac{a}{2}$.}\label{DMC} 
\end{figure}
\end{center}
\begin{align}
  G_{\textrm{BC}}(x,x^\prime)=\frac{2}{a}\int\frac{\textrm{d}\omega}{2\pi}~e^{\omega(t-t^\prime)}
                \sum\limits_n\frac{\textrm{sin}[k_n(x+\frac{a}{2})]\textrm{sin}[k_n(x^\prime+\frac{a}{2})]} {\omega^2+k_n^2+m^2},
\end{align}
where $a$ denotes the distance between the plates.
The Green\rq{}s function in the free or no boundary case is,
\begin{align}
G_{\textrm{free}}(x,x^\prime)=\int\frac{\textrm{d}^{2}k}{(2\pi)^2}\frac{ie^{-ik.(x-x^\prime)}}{k^2-m^2+i\epsilon}.
\end{align}
These two Green\rq{}s functions diverge as $x\to x^\prime$.
Using  Eq.\,(\ref{m.c}), the difference between the corresponding counter-terms is,
\begin{align}
\Delta\big[(\delta m(x)\big]=\delta m_{\textrm{BC}}-\delta m_{\textrm{free}}
=-\frac{\lambda}{2}\left[G_{{\textrm{BC}}}(x,x)-G_{\textrm{free}}(x,x)\right].
\end{align}
In Fig.\,(\ref{DMC}) we plot $\Delta\big[\delta m(x)\big]$.
Note that this quantity is extremely small, as compared to each 
of the divergent counter-terms, except at the boundaries. 
In general the value of $\Delta\big[\delta m(x)\big]$ is only significant 
for distances within a Compton wave-length from the end points.

\end{document}